\documentclass[11pt]{book}
\usepackage{amssymb}
\usepackage{amsmath}
\usepackage{graphicx}
\usepackage{subfigure}
\usepackage{makeidx}
\usepackage{multicol}
\begin{document}

\def\aj{\rm{AJ}}                   
\def\araa{\rm{ARA\&A}}             
\def\apj{\rm{ApJ}}                 
\def\apjl{\rm{ApJ}}     
\def\actaa{\rm{ACTAA}}         
\def\apjs{\rm{ApJS}}               
\def\ao{\rm{Appl.~Opt.}}           
\def\apss{\rm{Ap\&SS}}             
\def\aap{\rm{A\&A}}                
\def\aapr{\rm{A\&A~Rev.}}          
\def\aaps{\rm{A\&AS}}              
\def\azh{\rm{AZh}}                 
\def\baas{\rm{BAAS}}               
\def\jrasc{\rm{JRASC}}             
\def\memras{\rm{MmRAS}}            
\def\mnras{\rm{MNRAS}}             
\def\pra{\rm{Phys.~Rev.~A}}        
\def\prb{\rm{Phys.~Rev.~B}}        
\def\prc{\rm{Phys.~Rev.~C}}        
\def\prd{\rm{Phys.~Rev.~D}}        
\def\pre{\rm{Phys.~Rev.~E}}        
\def\prl{\rm{Phys.~Rev.~Lett.}}    
\def\pasp{\rm{PASP}}               
\def\pasj{\rm{PASJ}}               
\def\qjras{\rm{QJRAS}}             
\def\skytel{\rm{S\&T}}             
\def\solphys{\rm{Sol.~Phys.}}      
\def\sovast{\rm{Soviet~Ast.}}      
\def\ssr{\rm{Space~Sci.~Rev.}}     
\def\icarus{\rm{Icarus}}     
\def\zap{\rm{ZAp}}                 
\def\nat{\rm{Nature}}              
\def\iaucirc{\rm{IAU~Circ.}}
\def\aplett{\rm{Astrophys.~Lett.}}
\def\apspr{\rm{Astrophys.~Space~Phys.~Res.}}
\def\bain{\rm{Bull.~Astron.~Inst.~Netherlands}}
\def\fcp{\rm{Fund.~Cosmic~Phys.}}
\def\gca{\rm{Geochim.~Cosmochim.~Acta}}
\def\grl{\rm{Geophys.~Res.~Lett.}}
\def\jcp{\rm{J.~Chem.~Phys.}}      
\def\jgr{\rm{J.~Geophys.~Res.}}    
\def\jqsrt{\rm{J.~Quant.~Spec.~Radiat.~Transf.}}
\def\memsai{\rm{Mem.~Soc.~Astron.~Italiana}}
\def\nphysa{\rm{Nucl.~Phys.~A}}
\def\physrep{\rm{Phys.~Rep.}}
\def\physscr{\rm{Phys.~Scr}}
\def\planss{\rm{Planet.~Space~Sci.}}
\def\procspie{\rm{Proc.~SPIE}}
\let\astap=\aap
\let\apjlett=\apjl
\let\apjsupp=\apjs
\let\applopt=\ao


\chapter[Bloom \& Richards]{Data Mining and Machine-Learning in Time-Domain Discovery \& Classification}


{\Large \bf Joshua S. Bloom}\vspace{0.05in} \\{\it Astronomy Department\\University of California, Berkeley\\601 Campbell Hall, Berkeley, CA 94720}\\ {\tt jbloom@astro.berkeley.edu}\\

\noindent {\Large \bf Joseph W. Richards}\vspace{0.05in} \\{\it Astronomy Department/Statistics Department\\University of California, Berkeley\\ 601 Campbell Hall, Berkeley, CA 94720}\\ {\tt jwrichar@stat.berkeley.edu}\\

The changing heavens have played a central role in the scientific effort of astronomers for centuries. Galileo's synoptic observations of the moons of Jupiter and the phases of Venus starting in 1610, provided strong  refutation of Ptolemaic cosmology. These observations came soon after the discovery of Kepler's supernova had challenged the notion of an unchanging firmament.  In more modern times, the discovery of a relationship between period and luminosity in some pulsational variable stars \cite{1908AnHar..60...87L} led to the inference of the size of the Milky Way, the distance scale to the nearest galaxies, and the expansion of the Universe (see \cite{2010ARA&A..48..673F} for review). Distant explosions of supernovae were used to uncover the existence of dark energy and provide a precise numerical account of dark matter (e.g., \cite{2006A&A...447...31A}). Repeat observations of pulsars \cite{1992Natur.355..145W} and nearby main-sequence stars revealed the presence of the first extrasolar planets \cite{1995Natur.378..355M,1998ARA&A..36...57M,2000ApJ...529L..41H,2000ApJ...529L..45C}. Indeed, time-domain observations of transient events and variable stars, as a technique, influences a broad diversity of pursuits in the entire astronomy endeavor \cite{2009astro2010S.307W}.

\begin{figure}[tbh]
\centerline{\includegraphics[width=5.0in,angle=0]{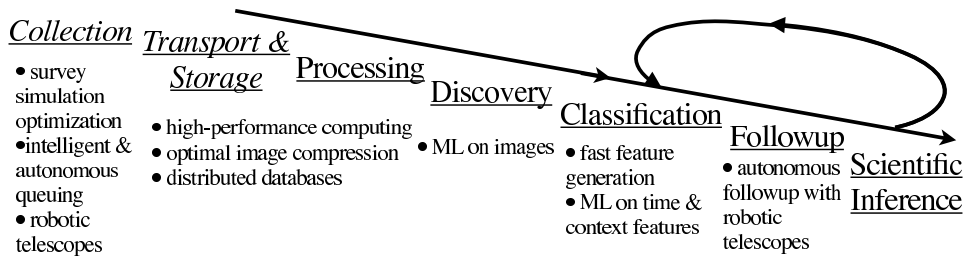}}
\caption[Data mining, computation, and machine learning roles in the scientific process.]{Data mining, computation, and ML roles in the scientific pathway.}
\label{fig:science}
\end{figure}

While, at a fundamental level, the nature of the scientific pursuit remains unchanged, the advent of astronomy as a {\it data-driven} discipline presents fundamental challenges to the way in which the scientific process must now be conducted.  Digital images (and data cubes) are not only getting larger, there are more of them. On logistical grounds, this taxes storage and transport systems. But it also implies that the intimate connection that astronomers have always enjoyed with their data---from collection to processing to analysis to inference---necessarily must evolve. Figure \ref{fig:science} highlights some of the ways that the pathway to scientific inference is now influenced (if not driven by) modern automation processes, computing, data-mining and machine learning. 

The emerging reliance on computation and machine learning (ML) is a general one---a central theme of this book---but the time-domain aspect of the data and the objects of interest presents some unique challenges. First, any collection, storage, transport, and computational framework for processing the streaming data must be able to keep up with the dataflow. 
This is not necessarily true, for instance, with static sky science, where metrics of interest can be computed off-line and on a timescale much longer than the time required to obtain the data. Second, many types of transient (one-off) events evolve quickly in time and require more observations to fully understand the nature of the events. This demands that time-changing events are quickly discovered, classified, and broadcast to other followup facilities. All of this must happen robustly with, in some cases, very limited data. Last, the process of discovery and classification must be calibrated to the available resources for computation and followup. That is, the {\it precision} of classification must be weighed against the {\it computational cost} of producing that level of precision. Likewise, the cost of being wrong about the classification of some sorts of sources must be balanced against the scientific gains about being right about the classification of other types of sources. Quantifying these tradeoffs, especially in the presence of a limited amount of followup resources (such as the availability of larger-telescope observations) is not straightforward and inheres domain-specific imperatives that will, in general, differ from astronomer to astronomer.

\begin{figure}[tbh]
\centerline{\includegraphics[width=4in]{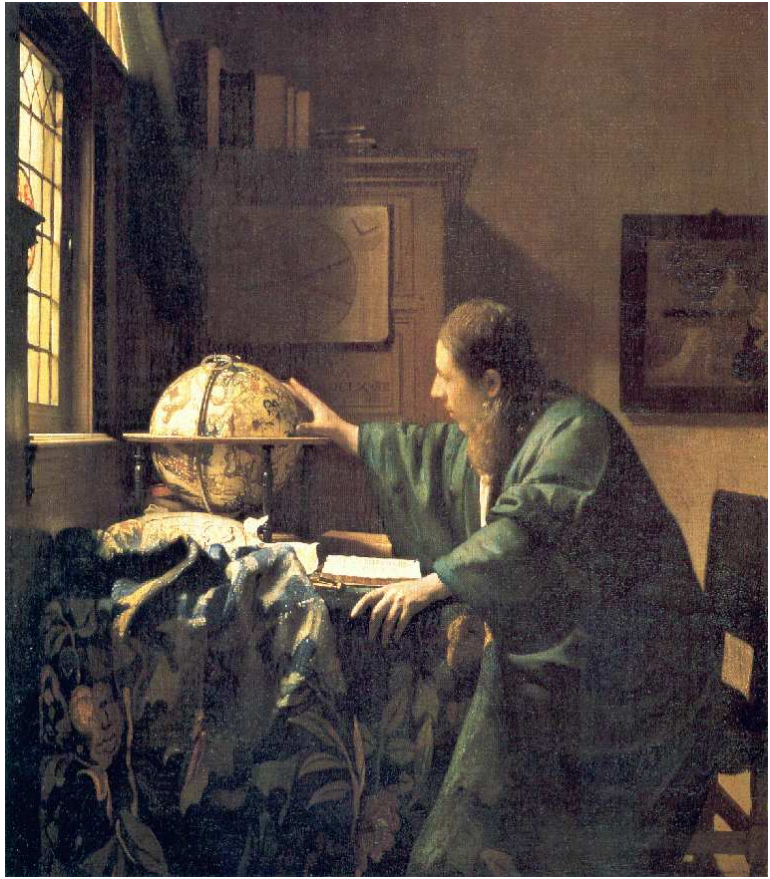}}
\caption[Vermeer's {\it The Astronomer}]{``The Astronomer,'' by Johannes Vermeer, c.~1668. In many ways, epoch of the armchair astronomer is returning to primacy.}
\label{fig:astronomer}
\end{figure}

This chapter presents an overview of the current directions in machine learning and data-mining techniques in the context of time-domain astronomy. Ultimately the goal---if not just the necessity given the data rates and the diversity of questions to be answered---is to abstract the traditional role of astronomer in the entire scientific process. In some sense, this takes us full-circle from the pre-modern view of the scientific pursuit presented in Vermeer's ``The Astronomer'' (Figure \ref{fig:astronomer}): in broad daylight, he contemplates the nighttime heavens from depictions presented to him on globe, based on observations that others have made. He is an abstract thinker, far removed from data collection and processing; his most visceral connection to the skies is  just the feel of the orb under his fingers. Substitute the globe for a plot on a screen generated from an SQL query to a massive public database in the cloud, and we have a picture of the modern astronomer benefitting from the ML and data-mining tools operating on an almost unfathomable amount of raw data.

\section{Discovery}

We take the notion of discovery, in the context of the time domain, as the recognition that data collected (e.g., a series of images of the sky) contains a source which is changing in time in some way. Classification (\S \ref{sec:class}) is the quantification of the similarity of that source to other known types of variability and, by extension, the inference of {\it why} that source is changing.
The most obvious change to discover is that of brightness or flux. On imaging data, changes in color and position might also be observed\footnote{Discovery of change in position, especially for fast-moving sources (such as asteroids), inheres its own set of data-mining challenges which we will discuss. See, for example, \cite{2007Icar..189..151K,2010PASP..122..549P}.}. Spectroscopically, changes in emission/absorption properties and apparent velocities might also be sought. Discovery of time-variable behavior is technique-specific and, as such, we will review the relevant regimes. Yip et al.~\cite{2009AJ....137.5120Y} discuss variability discovery on spectroscopy line features in the context of active galactic nuclei. Gregory \cite{2011MNRAS.410...94G} presents ML-based discovery and characterization algorithms for astrometric- and Doppler-based data in the context of exoplanets. We focus here on the discovery of brightness/flux variability.

\subsection{Identifying Candidates}

{\bf Pixelated Imaging}: Many new and planned wide-field surveys are drawing attention to the need for data-mining and ML.  These surveys will generate repeated images of the sky in some optical or infrared bandpass. These 2-dimensional digitized images form the basic input to discovery\footnote{In each one minute exposure, for example, the Palomar Transient Factory \cite{2010SPIE.7735E.122L} produces 11 images each from a 2k $\times$ 4k CCD array of size 1 sq.~arcsecond (0.65 sq.~degree per image). Since each pixel is 2 bytes, the amounts to 184 MB of raw data generated per minute.  Raw data are pre-processed using calibration data to correct for variable gain and illumination across the arrays; spatially-dependent defects in the arrays are flagged and such pixels are excluded from further scrutiny.}.  The data from such surveys are usually obtained in a ``background-limited'' regime, meaning that the signal-to-noise on an exposure is dominated by the flux of sources (as the signal) and the background sky brightness (as the dominant noise component). Except in the most crowded images of the plane of the Milky Way, most pixels in the processed images contain only sky flux. Less than a few percent of pixels usually contain significant flux from stars, galaxies or other astrophysical nebulosities.

There are two broad methods for discovering variability in such images. In one, all sources above some statistical threshold of the background noise are found and the position and flux associated with those sources are extracted to a catalog. There are off-the-shelf codebases to do this (e.g., \cite{1996A&AS..117..393B,2002ApJS..138..185F}) but such detection and extraction on images is by no means straightforward nor particularly rigorous, especially near the sky-noise floor of images. Discovery of variability is found by asking statistical questions (see \S \ref{sec:lc}) about the constancy (or otherwise) on the {\it light curve} produced on a given source, created by cross-correlating sources by their catalog position across different epochs \cite{2008ASPC..394..165B}. The other method, called ``image differencing,'' \cite{1996AJ....112.2872T,1998ApJ...503..325A,2000AcA....50..421W,2008MNRAS.386L..77B} takes a new image and subtracts away a ``reference image'' of the same portion of the sky; this reference image is generally a sharp, high signal-to-noise composite of many historical images taken with the same instrumental setup and is meant to represent an account of the ``static'' (unchanging) sky. 

Both methods have their relative advantages and drawbacks (see \cite{2009ApJ...696..870D} for a discussion). Since image differencing involves astrometric alignment and image convolution, catalog-based searches are generally considered to be faster. Moreover, catalog searches tend to produce fewer spuriously detected sources because the processed individual images tend to have less ``defects'' than differenced images. Catalog searches perform poorly, however, in crowded stellar fields (where aperture photometry is difficult) and in regions around galaxies (where new point sources embedded in galaxy light can be easily outshined).  Given the intellectual interests in finding variables in crowded fields (e.g., microlensing; \cite{1986ApJ...304....1P,2001MNRAS.327..868B}) and transient events (such as supernovae and novae) near galaxies, image-difference based discovery is considered necessary for modern surveys.

Computational costs aside, one of the primary difficulties with image differencing is the potential for a high ratio of spurious candidate events to truly astrophysical events. A trained human scanner can often discern good and bad subtractions and, for many highly successful projects, human scanners were routinely used for determining promising discovery candidates. The KAIT supernova search \cite{2001ASPC..246..121F} makes use of undergraduate scanners to shift through $\sim1000$ images from the previous night. Over 1000 SNe were discovered in 10 years of operations with this methodology \cite{2010arXiv1006.4611L}. Basic quality/threshold cuts on the metrics about each candidate can be used to present to human scanners a smaller subset of images for inspection; in this way, the Sloan Digital Sky Survey II Supernova Search \cite{2008AJ....135..338F} netted $>300$ spectroscopically confirmed supernovae discoveries from $\sim$150,000 manually scanned candidates. The Nearby Supernova Factory \cite{2002SPIE.4836...61A}, after years of using threshold cuts, began to use boosted decision trees (\S \ref{sec:super}) on metrics from image differences to optimize supernova discovery \cite{2007ApJ...665.1246B}.
Unlike with specific domain-focused discovery surveys (like supernovae searches), many surveys are concerned with discovery and classification of all sorts of variable stars and transients. So unlike in the supernova discovery classifier of Bailey et al.~\cite{2007ApJ...665.1246B} (which was highly tuned to finding transient events near galaxies), discovery techniques must aim to be agnostic to the physical origin of the source of variability. That is, there is an imperative to separate the notion of ``discovery'' and ``physical classification.''  

\begin{figure}[tbp]
\centerline{\includegraphics[width=200pt,angle=270]{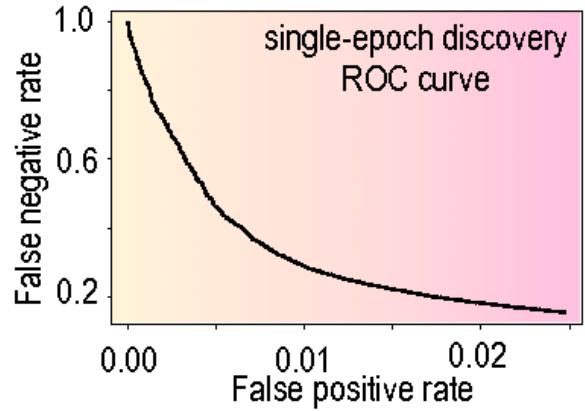}}
\caption[ROC Curve for Image-Differenced Candidates]{ROC curve for image-differenced candidates based on a training sample of 50,000 candidates from the Palomar Transient Factory. High efficiency and high purity are on the bottom left of the plot. From \cite{sahand2011}.}
\label{fig:rfcandidate}
\end{figure}

In the Palomar Transient Factory, we find at least one hundred high-significance bogus candidates for every one real candidate in image differences \cite{bloom2011}. With over one million candidates produced nightly, the number of images that would have to be vetted by humans is unfeasible. Instead, we produced a training set of human-vetted candidates, each with dozens of measured features (such as FWHM, ellipticity; see  \cite{bloom2011}). These candidates are scored on a scale from 0 to 1 based on their inferred likelihood of being bogus or astrophysically ``real.'' We developed a random forest classifier on the features to predict the 1--0 real-bogus value and saved the result of the ML-classifer on each candidate.  These results are used to make discovery decisions in PTF. After one year of the survey, we also created training sets of real and bogus candidates by using candidates associated with known/confirmed transients and variables \cite{sahand2011}. Figure \ref{fig:rfcandidate} shows the ``receiver operating characteristic'' (ROC) curve for a random forest classifier making use of the year-one training sample.

If all real sources occurred at just one epoch of observation, then ROC curves such as those depicted in Figure \ref{fig:rfcandidate} would directly reflect discovery capabilities: type I error (false-negatives) would be the efficiency for discovery and type II error (false-positive rate) would be the purity for discovery. However, most transient events occur over several epochs and bogus candidates often do not recur at precisely the same location.  Therefore, turning candidate-level ROC curves to global discovery efficiency/purity quantities is not straightforward. In PTF we require 2 high-quality ML-score candidates within a 12 day window to qualify a certain position on the sky as a discovery of a true astrophysical source\footnote{This discovery is designed to find fast-changing events, of particular interest to the PTF collaboration. We also require at least two observations more than 45 minutes separated in time, to help remove moving asteroids from the discovery set.}.  In the first 8 months of producing automatic discoveries with PTF, our codebase independently discovered over 10,000 transients and variable stars.

{\bf Radio Interferometry}: Traditionally, radio images are generated from raw u-v interferometric data using a human intensive process to iteratively flag and remove spurious baseline data. Phase drift due to the ionosphere, instrumental instability, and terrestrial radio-frequency interference (RFI) are all impediments to automatically producing clean images of the sky. Given the massive data rates soon expected from wide-field surveys (e.g., LOw Frequency ARray: LOFAR; Australian Square Kilometre Array Pathfinder; ASKAP), there is a pressing need to autonomously produce clean images of the radio sky. Algorithmic innovations to speed  automatic image creation have been impressive (e.g., \cite{2004SPIE.5489..817N}). For RFI mitigation, a genetic algorithm approach has produced promising results \cite{2005RaSc...40S5S08F}. Once images are made, sources are detected much the same way as with optical imaging\footnote{Note that McGowan et al.~\cite{2005ASPC..345..362M} have developed an ML approach to faint source discovery in radio images.} and catalog searches are used to find transients and variables \cite{2007ApJ...666..346B,2010ApJ...719...45C}.

\subsection{Detection and Analysis of Variability}
\label{sec:lc}

For catalog-based searches, variability is determined on the basis of the collection of flux measurement as a function of time for a candidate source.  Since variability can be manifested in many ways (such as aperiodic behavior, occasional eclipsing etc.) one single metric on variability will not suffice in capturing variability \cite{1996PASP..108..851S,2005ESASP.576..513E,2008MNRAS.386..887B,2009MNRAS.400.1897S,2009BlgAJ..12...49D}.  A series of statistical questions can be asked with each new epoch to each light curve. Are the data consistent with an unchanging flux, in a $\chi^2$ sense? Are there statistically significant deviant data points? How are those outliers clustered in time?  Significant variability of periodic sources may be revealed by direct periodogram analysis (\cite{1989MNRAS.241..153S}; see also ref.\ \cite{2011arXiv1101.2445B}). In the Poisson detection limit, such as at $\gamma$-ray wavebands or with detections of high-energy neutrinos, discovering variability in a source is akin to asking the question of whether there is a statistically significant change in the rate of arrival of individual events; for this, there are sophisticated tools (such as Bayesian blocks) for analysis \cite{1998ApJ...504..405S,2001ApJ...550L.101S}. One of the important real-world considerations is that photometric uncertainty estimates are always just estimates, based on statistical sampling of individual image characteristics. Systematic errors in this uncertainty (either too high or too low) can severely bias variability metrics (c.f.\ \cite{2005ESASP.576..513E}).  Characterizing efficiency-purity from systematic errors must be done on a survey by survey basis. 

\section{Classification}
\label{sec:class}

Determining the physical origin of variability is the basic impetus of classification. But clearly what is {\it observed} and what is {\it inferred to belie that which is observed} are not the same, the latter deriving from potentially several interconnected and complex physical processes. A purely physical-based classification schema is then reliant upon subjective and potentially incorrect model interpretation. For instance, to say that the origin of variability is due to an eclipse  requires an intuitive leap, however physically relevant, from observations of a periodic dip in an otherwise constant light curve. A purely observational-based classification scheme, on the other hand, lacks the clarifying simplicity offered by physical classification. For example, how is a periodic light curve ``dipping'' (from an eclipsing system) different, quantitatively, than an extreme example of periodic brightness changes (say from a pulsational variable)? To this end, existing classification taxonomies tend to rely on an admixture of observational and physical statements.  And when a variable source is found the goal is in finding how that source fits within an established taxonomy.

Phenomenological and theoretical taxonomies aside, the overriding conceptual challenge of classification is that no two sources in nature are identical and so the boundaries between classes (and subclasses) are inherently fuzzy: there is no ground truth in classification, regardless of the amount and quality of the data. With finite data, the logistical challenge is in extracting the most relevant information, mapping that onto the quantifiable properties derivable from instances of other variables, and finding an (abstractly construed) distance to other sources. There a several broad reasons for classification:
\begin{enumerate}
	\item {\bf Physical Interest} Understanding the physical processes behind the diversity of variability requires numerous examples across the taxonomy. Studying the power-spectrum of variability in high signal-to-noise light curves can be used to infer the interior structure of stars (astroseismology). Modelling detached eclipsing systems can be used to infer the mass, radius, and temperatures of the binary components.
	\item {\bf Utility} Many classes of variables have direct utility in making astrophysically important measurements that are wholly disconnected from the origin of the variability itself. Mira, RR Lyrae, and Cepheids are used for  distant ladder measurements, providing probes of the structure and size of the universe. Calibrated standard-candle measurements of Ia and IIP supernovae are cosmographic probes of fundamental parameters. Short period AM CVn systems serve as a strong source of ``noise'' for space-based gravity wave detectors; finding and characterizing these systems through optical variability allows the sources to be  effectively cleaned out of the LISA datastream, allowing more sensitivity searches for gravity waves in the same frequency band.
	\item {\bf Demographics} Accounting for various biases, the demographics from classification of a large number of variable stars can be used to form and understand the evolutionary life-cycle of stars across mass and metallicity. Understanding the various ways in which high mass stars die can be gleaned from the demographics of supernova sub-types. 
	\item {\bf Rarities and Anomalies} Finding extreme examples of objects from known classes or new examples of sparsely populated classes has the potential to inform the understanding of (the more mundane) similar objects. The ability to identify anomalous systems and discover  new types of variables---either hypothesized theoretically or not---is likewise an important feature of any classification system.
\end{enumerate}

Expert-based (human) classification has been the traditional approach to time-series classification: a light-curve (and colors, and position on the sky, etc.) is examined and a judgement is made about class membership.  The preponderance of peculiar outliers of one (historical) class may lead to a consensus that a new sub-class is warranted\footnote{For example, type Ia supernovae, likely due to the explosion of a white dwarf, appear qualitatively similar in their light curves to some core-collapsed supernovae from hydrogen-stripped massive stars (Type Ib/Ic). Yet the presence or absence of silicon in the spectra became the defining observation that led to very different physical inferences for similar phenomenological types of supernovae.}. Again, with surveys of hundreds of thousands to billions of stars and transients,  this traditional role must necessarily be replaced by ML and other data-mining techniques. 

\subsection{Domain-based Classification}

\begin{figure}[tbp]
\centerline{\includegraphics[width=3.6in,angle=0]{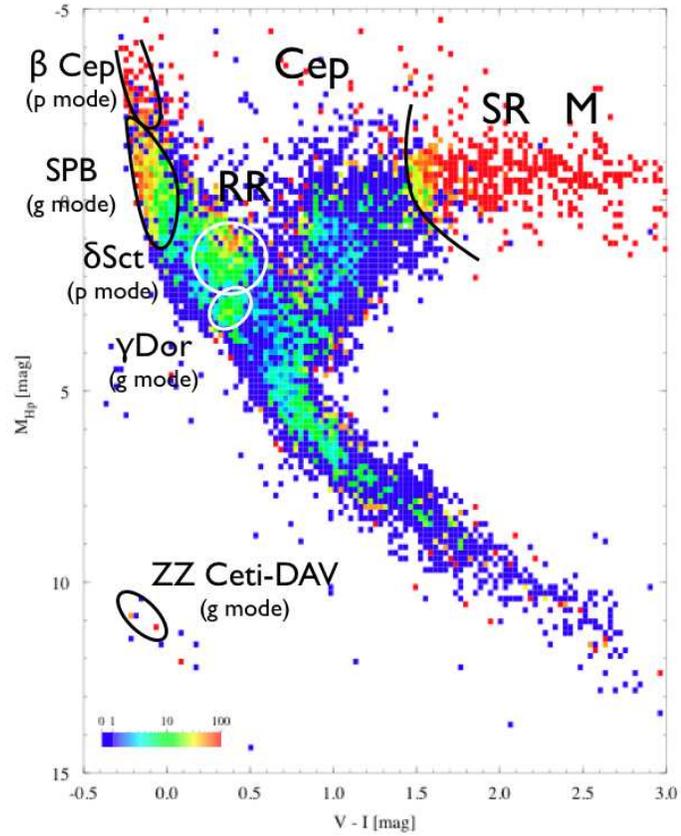}}
\caption[H-R diagram of variable stars]{Fractional variability of stars across the H-R diagram derived from Hipparcos data. Red indicates significant variability and blue low-amplitude variability ($10$\% peak-to-peak). Identification of colors coupled with distances provide a rather clean path to classification. From \cite{em08}.}
\label{fig:hr}
\end{figure}

\begin{figure}[tbp]
\centerline{\includegraphics[width=4.5in,angle=0]{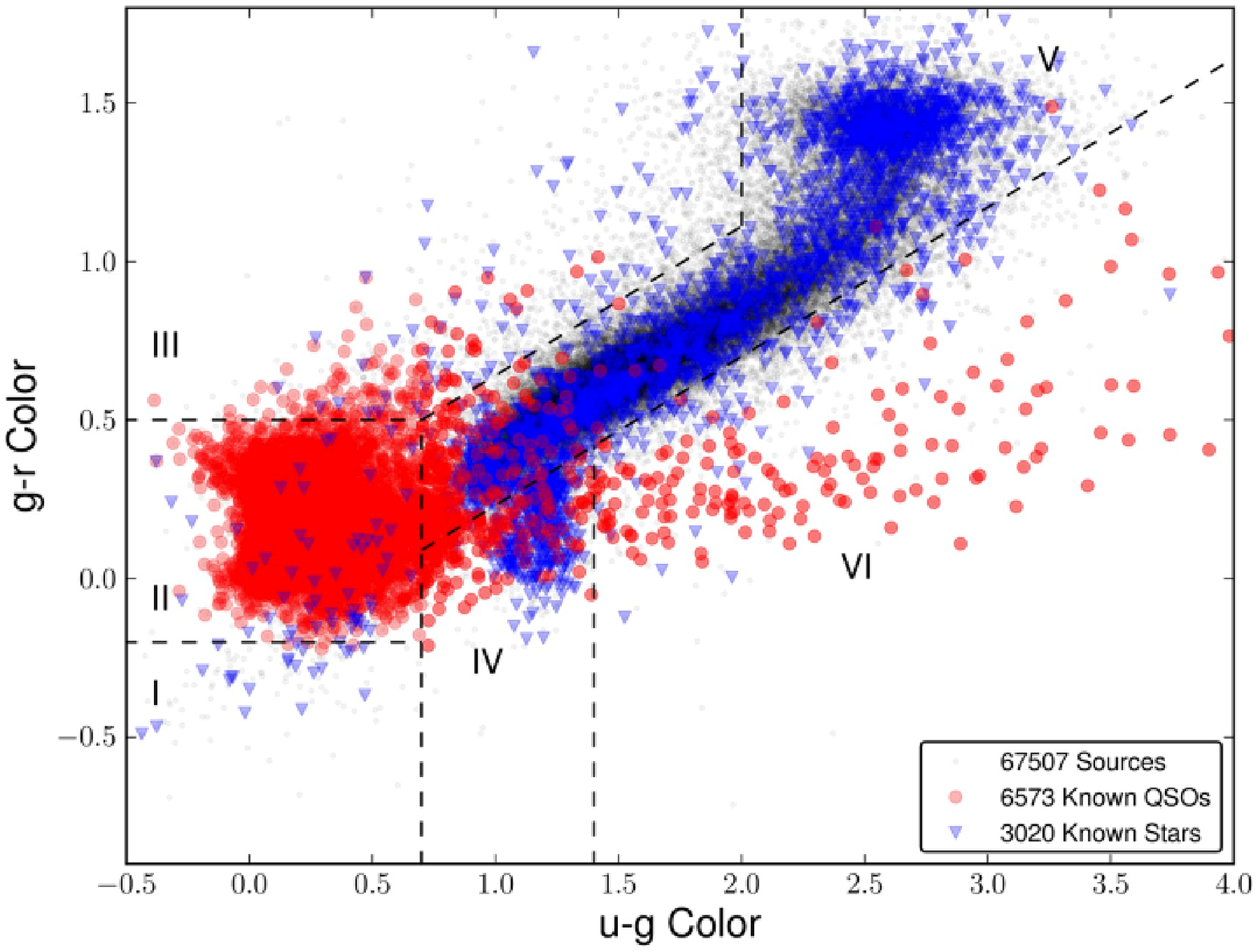}}
\caption[Color-color plot]{Color-color plot showing variable sources from Stripe 82. Region II is the traditional QSO locus and Region IV is the region populated by most RR Lyrae. There are clearly many QSOs that fall outside region IV (particularly high redshift QSOs), some of which are in the RR Lyrae region. From \cite{2010arXiv1008.3143B}.}
\label{fig:color}
\end{figure}

Some of the most fruitful modern approaches to classification involve domain-specific classification: using theoretical and/or empirical models of certain classes of interest to determine membership of new variables in that class. Once a source is identified as variable, its location in color-luminosity space can often provide overwhelming evidence of class membership (Figure \ref{fig:hr}). Hertzsprung-Russell (H-R) diagrams obviously require distance to the source to be known accurately and so, until Gaia \cite{2002Ap&SS.280....1P}, it has its utility restricted to those with parallax previously measured by the Hipparcos survey. For some sources, such as RR Lyrae and quasars, location in color-color space suffices to provide probable classification (Figure \ref{fig:color}). Strict color cuts or more general probabilistic decisions on clustering\footnote{Such classification decisions can make use of the empirical distribution of sources within a class and uncertainties on the data for a given instance \cite{2010arXiv1011.6392B}.} within a certain color-color space can performed. (Regardless, reddening and contamination often make such classification both inefficient and impure.)

Of considerable interest, given that historical and/or simultaneous color information is not always available and that unknown dust can affect color-based classification, is to classify using time-series data alone. For some domains, the light curves tell much of the story. Well before the peak brightness in a microlensing event, for example, an otherwise quiescent star will appear to brighten monotonically like a second-order power-law in time. By continuously fitting the light curve of (apparently) newly variable stars for such a functional form, a statistically rigorous question can be asked about whether that event appears to be microlensing or not. For a sufficiently-homogeneous class of variables, an empirical light curve can be fit to the data and those sources with acceptable fits can be admitted to that class. This was done to discover and classify RR Lyrae stars in the SDSS Stripe 82 dataset \cite{2010ApJ...708..717S}.  Such approaches require, implicitly, a threshold of acceptability.  However, using cuts based on model probabilities and goodness-of-fit values can be damaging:  these metrics are often a poor description of class probabilities due to the overly-restricted space of template models under consideration as well as other modeling over-simplifications.  A better approach is to use a representative training set of sources with known class to estimate the ROC curve for the model fits, and to then pick the threshold value corresponding with the desired efficiency and purity of the sample.  If the training set is truly representative, this ensures a statistical guarantee of the class efficiency and purity of samples generated by this approach.

\begin{figure}[tbh]
\centerline{\includegraphics[width=4.8in,angle=0]{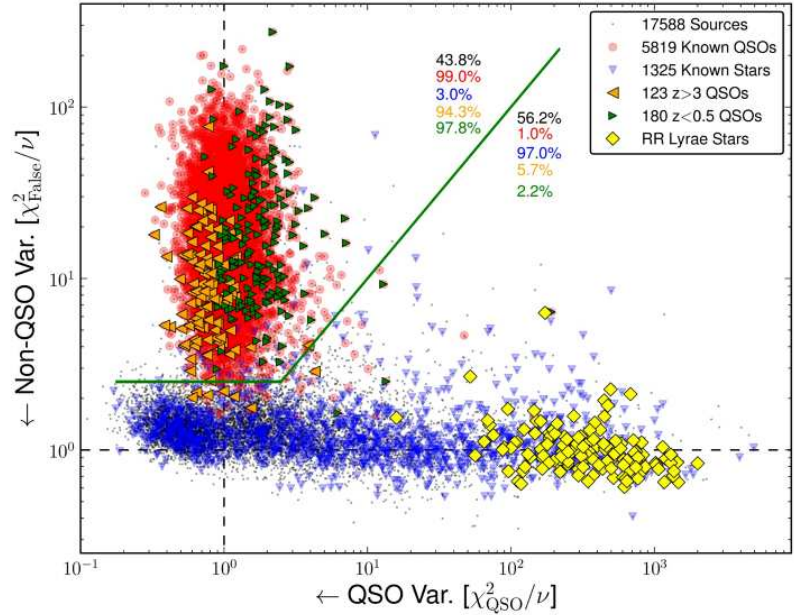}}
\caption[QSO Variability]{Variability selection of quasars. Using a Bayesian framework to connect light-curves of point sources to damped random walk behavior, statistics that account for uncertainty and covariance can be developed to find QSO-like behavior. This selection (green line) is highly efficient at finding known QSOs ($\sim$99\%) and impure at the $3$\% level. From \cite{2010arXiv1008.3143B}.}
\label{fig:qso}
\end{figure}

A related, but less strong statement can often be made that the variability has ``class-like'' variability. For example, there is no one template of a quasar light curve but since quasars are known to vary stochastically like a damped random walk, with some characteristic timescale that correlates only mildly with luminosity, it is possible to capture the notion of whether a given light curve is statistically consistent with such  behavior. In Butler \& Bloom \cite{2010arXiv1008.3143B} we created a set of features designed to capture how much a variable was ``quasar like'' and found a high degree of efficiency and purity of quasar identification based on a spectroscopic validation sample (Figure \ref{fig:qso}). Some variable stars, such pulsating super giants and X-ray binaries, also show this QSO-like behavior; so it is clear that such domain-specific statistical features alone cannot entirely separate classes.

There is significant utility in restricting the model fits to a finite number of classes. Indeed, one of the more active areas of domain-specific classification is in supernova subclassing. By assuming that a source is some sort of supernovae, a large library of well-observed supernova light curves (and photometric colors) can be used to infer the sub-type of a certain instance, especially when quantifying the light curve trajectory through color-color space \cite{2002PASP..114..833P}. Provided that the library of events spans (and samples) sufficiently well the space of possible subclasses (and making use of available redshift information to transform templates appropriately), Bayesian odds ratios can be effectively used to determine membership within calibrated confidence levels (see ref.~\cite{2010arXiv1010.1005N}).

\subsection{Feature-based Classification}

An abstraction from domain-specific classification (such as template fitting) is to admit that the totality of the available data belies the true classification, irrespective of whether we understand the origin of that variability or have quantified specifically what it means to belong to a certain class. We classify on {\it features}, metrics derived from time-series and contextual data. There are a number of practical advantages to this transformation of the data. First, feature creation allows heterogeneous data to be mapped to a more homogeneous $m$-dimension real number line space. In this space, instances of variable objects collected from different instruments with different cadences and sensitivities can be directly intercompared. This is the sort of space where machine-learning algorithms work well, allowing us to bring to bear the richness of the machine-learning literature to astronomical classification. Second, features may be arbitrary simple (e.g., median of the data) or complex. So in cases with only limited data availability---when, for instance, light curve fitting might fail---we have a subset of metrics that can still be useful in classification. Many machine-learning frameworks have prescriptions for dealing with missing data that do not bias the results.  Third, many feature-based classification methods produce class probabilities for each new source, and there are well-prescribed methods in ML both for calibrating the classification results and to avoiding overfitting.  Last, ML approaches allow us to explicitly encode the notion of loss (or ``cost'') in the classification process, allowing for a controlled approach to setting the efficiency and purity of the final results.

There is, of course, a huge space of possible features and many will be significantly related to others (e.g., mean and median will strongly correlate). One of the interesting advantages of some ML techniques is the classification robustness both in the face of feature covariance and ``useless'' features. This is freeing, at some level, allowing us to create many feature generators without worry that too many kitchen sinks will sink the boat. The flip side, however, is that there are always more features on the horizon than those in hand that could be incrementally more informative for a certain classification task.

Methods for feature-based classification of time-varying sources in astronomy come in one of two flavors.  The first are {\bf supervised} methods, which use both the features and previously-known class labels from a set of training data to learn a mapping from feature to class space.  The second are {\bf unsupervised} methods (also called statistical clustering), which do not use class labels and instead seek to unveil clustering of the data in feature space.  The end goals of these approaches are different:  supervised classification attempts to build an accurate \emph{predictive} model where, for new instances, the true classes (or class probabilities) can be predicted with as few errors as possible, whereas unsupervised classification seeks a characterization of the distribution of features, such as estimating the number of groups and allocation of the data points into those groups.  A common technique (e.g., \cite{2005MNRAS.358...30E}) is to blend the two by first performing unsupervised classification and subsequently analyzing the resultant clusters with respect to a previously-known set of class labels.

\subsubsection{Feature Creation}

The two broad classes of features, {\bf time-domain} and {\bf context}, each provide unique value to classification but also inhere unique challenges. The most straightforwardly calculated time-domain features are based on the distribution of detected fluxes, such as the various moments of the data (mean, skewness, kurtosis). Variability metrics, such as $\chi^2$ under an unchanging brightness hypothesis and the so-called Stetson variability quantities \cite{1996PASP..108..851S}, are easily derived and make use of photometric uncertainties. Quantile-based measurements (such as the fraction of data observed between certain flux ranges) provide some robustness to outliers and provide a different view of the brightness distribution than moments. Inter-comparisons (e.g., ratios) of these metrics across different filters may themselves be useful metrics. 

\begin{figure}[tbp]
\centerline{\includegraphics[width=1.8in,angle=270]{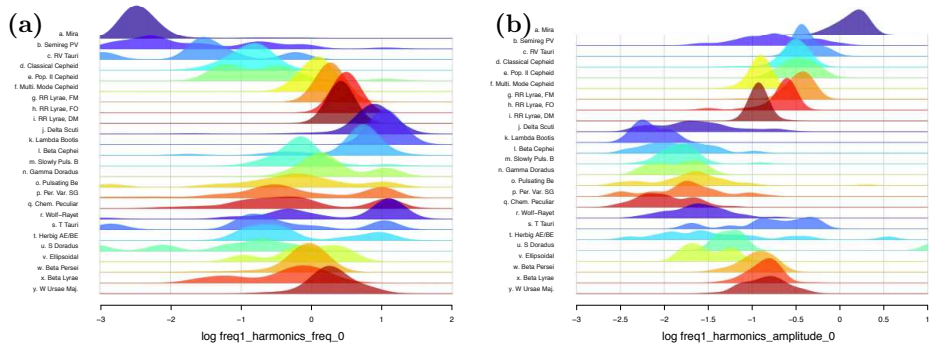}}
\caption[Distribution of features of variable stars]{Distribution of two frequency-domain features derived for 25 classes of variable stars from OGLE and Hipparcos photometry: a) log of the most significant frequency (units of day$^{-1}$) from a generalized Lomb-Scargle periodogram analysis, and b) the log of the amplitude of the most significant period, in units of magnitude. Mira variables (top) are long-period, high-amplitude variables, while delta Scuti stars (10th from top) are short-period, low-amplitude variables. Aperiodic sources, such as S Doradus stars (5th from bottom), have a large range in effective dominant period. From \cite{2011rich}.}
\label{fig:varstarfeature}
\end{figure}

Time-ordered metrics retain phase information. Frequency analysis, finding significant periodicity in the data, provides powerful input to the classification of variable stars (Figure \ref{fig:varstarfeature}). There are significant limitations to frequency-domain features, most obvious of which is that a lot of time-series data is required to make meaningful statements: with three epochs of data, it makes no sense to ask what the period of the source is. Even in the limit that a frequency of interest ($f_0$) is potentially sampled well in a Nyquist sense (where the total time duration of the light curve is longer than $\sim2/f_0$), the particular cadence of the observations may strongly alias the analysis, rendering significance measurements on peaks in the periodogram intractable. And unless the sources are regularly sampled (which, in general, they are not) there will be  covariance across the power spectrum. Finding significant periods can mean fitting a small amount of data over millions of trial frequencies, resulting in frequency-domain features that are computationally expensive\footnote{One practical approach for data observed with similar cadences is to compute the periodogram at a small number of a fixed set of frequencies and set the power/significance at each of these frequencies to be separate features. Covariance is then implicitly dealt with at the ML level, rather than feature generation level (e.g., ref.~\cite{2005MNRAS.358...30E})}. We review techniques and hybrid prescriptions for period finding and analysis in Richards et al.~\cite{2011rich}.

\begin{figure}[tbp]
\centerline{\includegraphics[width=3.2in,angle=270]{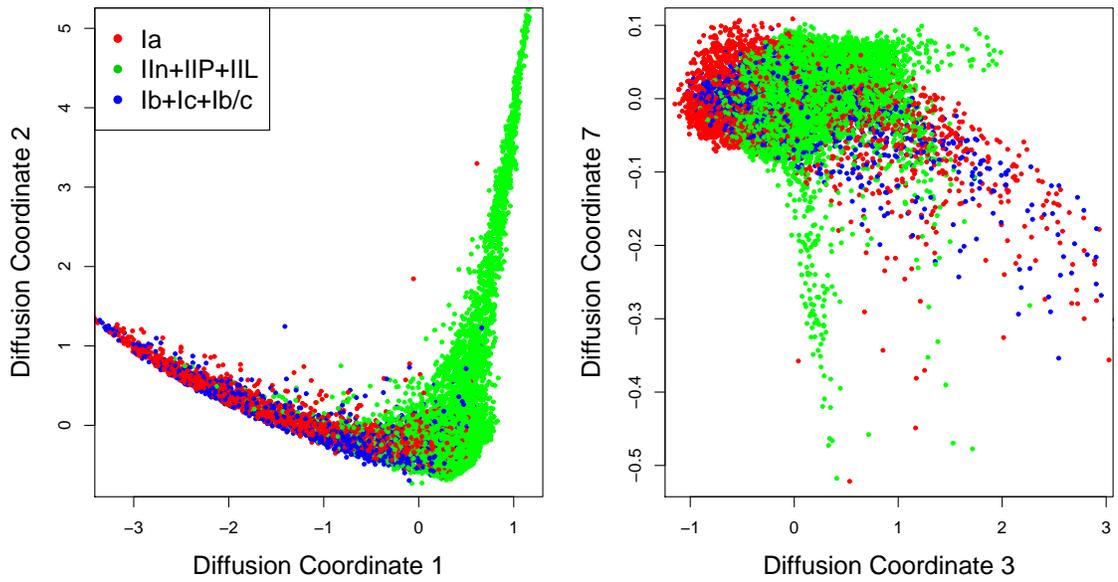}}
\caption[Diffusion map features of SNe]{Light curve distance measures can be used in conjunction with spectral methods, such as diffusion map, to compute informative features.  In this example, a spline-based distance between supernova light curves, designed to capture both shape and color differences, was used.  In the first two diffusion map coordinates (left), Type Ia and II SNe are distinguished, whereas higher features (right) reveal some separation between Ia and Ib/c supernovae. From \cite{2011richSN}.}
\label{fig:sndmap}
\end{figure}

Other time-ordered features may be extracted using a notion of ``distance'' between a given instance of a light curve and all others.  For instance, to derive features useful for supernova typing, Richards et al.~\cite{2011richSN} built up a matrix of pairwise distances between each pair of SNe (including both labeled and unlabeled instances) based on interpolating spline fits to the time-series measurements in each photometric band.  The pairwise distance matrix was subsequently fed into a diffusion map algorithm that embeds the set of supernovae in an optimal, low-dimensional feature space, separating out the various SN subtypes (Figure \ref{fig:sndmap}).  In a variable star analysis, Deb \& Singh~\cite{2009A&A...507.1729D} use the covariance matrix of a set of interpolated, folded light curves to find features using PCA.  In addition to using distance-based features to capture the time variability of sources, the way in which flux changes in time can be captured by fitting parameters under the assumption that the data are due to a Gaussian process \cite{springerlink:10.1007/978-3-540-28650-9_4}.

We define context-specific features as being all derivable features that are not expected to change in time. The location of the event on the sky, in Galactic or ecliptic coordinates, obviously provides a strong indication of whether the event has occurred in the Galaxy or in the Solar System. Metrics on the distance to the nearest detected galaxy and the parameters of that galaxy (its color, size, inferred redshift, etc.) are crucial features for determining the nature of extragalactic events. Even with very little time-domain data a strong classification statement can be made: for example, an event well off the ecliptic plane that occurs on the apparent outskirts of a red, spiral-less galaxy is almost certainly a type Ia supernova. One of the main challenges with context features is the heterogeneity of the available data. For example, in some places on the sky, particularly in the SDSS footprint, much is known about the stars and galaxies near any given position. Outside such footprints, context information may be much more limited. From a practical standpoint, if context information is stored only in remotely queryable databases, what information is  available  and the time it takes to retrieve that information may be highly variable in time. This can seriously affect the computation time to produce a classification statement on a given place on the sky.

\subsubsection{Supervised Approaches}
\label{sec:super}

\begin{figure}[tbp]
\centerline{\includegraphics[width=4in,angle=90]{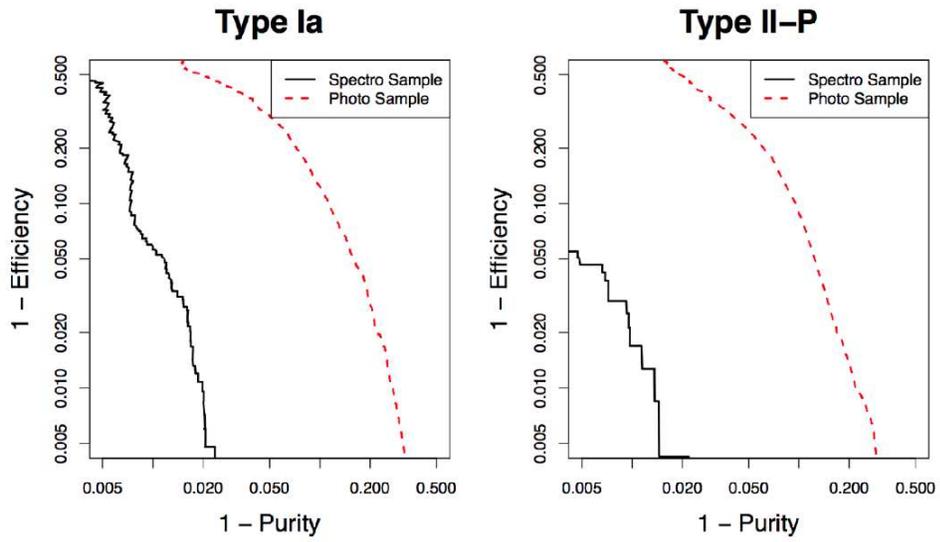}}
\caption[ROC curves for SN classification]{ROC curves for the selection of supernovae from a random forest probabilistic classifier, using data from the SN Photometric Classification Challenge \cite{2010PASP..122.1415K}.  Left: For classification of Type Ia SNe, in the spectroscopic sample we can achieve 95\% efficiency at a 99\% purity or $>99\%$ efficiency at 98\% purity, depending on the threshold.  Right: For Type II-P supernovae, the classifier performs even better, with higher efficiency at each given purity level. From \cite{2011richSN}.}
\label{fig:sneroc}
\end{figure}

Using a sample of light curves whose true class membership is known (e.g., via spectral confirmation), supervised classification methods learn a statistical model (known as a classifier) to predict the class of each newly-observed light curve from its features.  These methods are constructed to maximize the predictive accuracy of the classifications of new sources.  The goal of these approaches is clear: given a set of previously-labeled variables, make the best guess of the label of each new source (and optionally find the sources that do not fit within the given label taxonomy).   Many supervised classification methods also predict a vector of class probabilities for each new source.  These probabilistic classifiers can be used to compute ROC curves for the selection of objects from a specified science class---such as those in Figure \ref{fig:sneroc}---from which the optimal probability threshold can be chosen to create samples with desired purity and efficiency.

There are a countless number of classification methods in statistics and machine learning literature.  Our goal here is to review a few methods that are commonly used for supervised classification of time-variable sources in astronomy.

If the class-wise distributions of features were all completely known (along with the class prior proportions), then for a new source we would use Bayes' rule to compute the exact probability that the source is from each class, and classify the source as belonging to the class of maximal probability.  This is referred to as \emph{Bayes' classifier}, and is the provable best possible classifier in terms of error rate.  In practice, however, we do not know the class-wise feature distributions perfectly.  Many methods attempt to estimate the class densities from the training data.  In {\bf Kernel Density Estimation} (KDE) classification, the class-wise feature distributions are estimated using a non-parametric kernel smoother.  This approach has been used  to classify supernova light curves  \cite{2010arXiv1010.1005N}.  A pitfall of this technique is the tremendous difficulty in estimating accurate densities in high-dimensional feature spaces via non-parametric methods (this is referred to as the \emph{curse of dimensionality}).  To circumvent this problem, {\bf Na\"ive Bayes} performs class-wise KDE on one feature at a time, assuming zero covariance between features.  Though this simplifying assumption is unlikely to be true, Na\"ive Bayes has enjoyed much use, including in time-domain science \cite{2008AIPC.1082..287M}.  A step up from Na\"ive Bayes is {\bf Bayesian Network} classification, which assumes a sparse, graphical conditional dependence structure amongst the features.  This approach was used with considerable success  for variable star classification \cite{2007A&A...475.1159D,2009A&A...506..519D}.

Alternatively, class-wise distributions can be estimated using parametric models.  The {\bf Gaussian Mixture} classifier assumes that the feature distribution from each class follows a multivariate Gaussian distribution, where the mean and covariance of each distribution are estimated from the training data.  This approach is used widely in variable star classification (e.g., \cite{2007A&A...475.1159D}, \cite{2009A&A...506..519D}, and \cite{2010ApJ...713L.204B}).  The advantage of this parametric approach is that it does not suffer from curse of dimensionality.  However, if the data do not really follow a mixture of multivariate Gaussian distributions, then predictions may be inaccurate: for example, we showed in \cite{2011rich} that using the same set of variable star features, a random forest classifier outperforms the Gaussian mixture classifier by a statistically significant margin.  Gaussian mixture classifiers are also called {\bf Quadratic Discriminant Analysis} (QDA) classifiers (or {\bf Linear Discriminant Analysis}, LDA, if pooled covariance estimates are used).  These names refer to the type of boundaries that are induced between classes, in feature space.

\begin{figure}
\begin{center}
\includegraphics[angle=0,scale=.4]{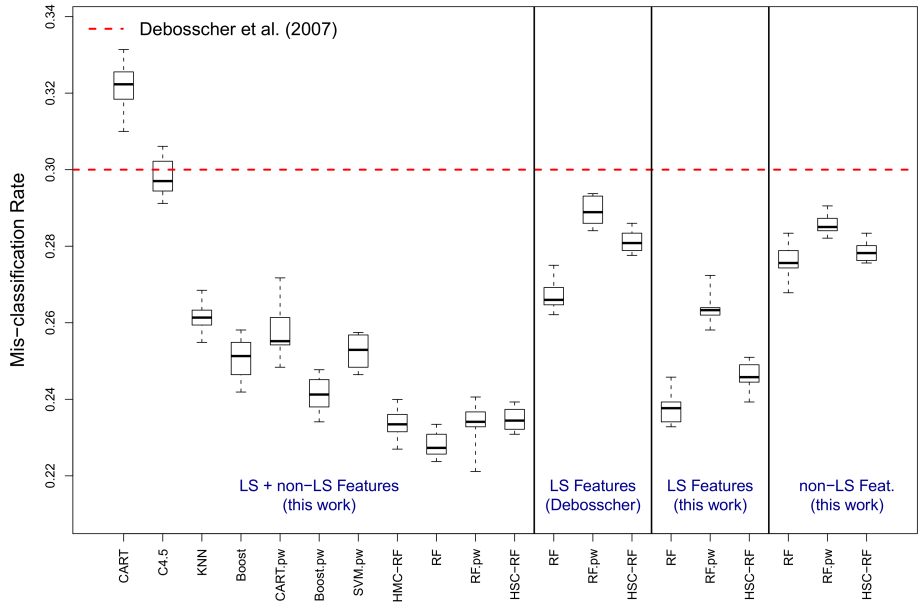}
\end{center}
\caption[Error rates for different classifers]{Distribution of cross-validation error rates for several classifiers on a mixed data set of OGLE and Hipparcos sources (see \cite{2011rich}).  The classifiers are divided based on the features on which they were trained; from left to right: (1) periodic plus non-periodic features, (2) the Lomb-Scargle features estimated by \cite{2007A&A...475.1159D}, (3) the Lomb-Scargle features estimated by \cite{2011rich}, and (4) only non-periodic features.  In terms of mis-classification rate, the random forest classifier trained on all of the features perform the best.  Classifiers considered are: classification trees (CART \& C4.5 variants), K-nearest neighbors (KNN), tree boosting (Boost), random forest (RF), pairwise versions of CART (CART.pw), random forest (RF.pw), and boosting (Boost.pw), pairwise SVM (SVM.pw), and two hierarchical random forest classifiers (HSC-RF, HMC-RF).  All of the classifiers plotted, except single trees, achieve better error rates than the best classifier from \cite{2007A&A...475.1159D} (dashed line), who considered Bayesian Network, Gaussian Mixture, ANN, and SVM classifiers. \label{fig:misclass}}
\end{figure}

Indeed, many classification methods instead focus on locating the optimal class boundaries.  {\bf Support Vector Machines} (SVMs) find the maximum-margin hyperplane to separate instances of each pair of classes.  Kernelization of a SVM can easily be applied to find non-linear class boundaries.  This is approach used to classify variable stars in a number of recent papers \cite{2007A&A...475.1159D,2007arXiv0712.2898W,2011rich}.   The {\bf K-nearest neighbors} (KNN) classifier predicts the class of each object by voting its K nearest neighbors in feature space, thereby implicitly estimating the class decision boundaries non-parametrically.  Another popular method is {\bf Classification Trees}, which performs recursive binary partitioning of the feature space to arrive at a set of pure, disjoint regions.  Trees are powerful classifiers because they can capture complicated class boundaries, are robust to outliers,  are immune to irrelevant features, and easily cope with missing feature values.  Their drawback is that due to their hierarchical nature, they tend to have high variance with respect to the training set.  Tree ensemble methods, such as {\bf Bagging}, {\bf Boosting}, and {\bf Random Forest} overcome this limitation by building many classification trees to bootstrapped versions of the training data and averaging their results.  Boosting, which has been used by Newling et al.~\cite{2010arXiv1010.1005N} for SN classification and Richards et al.~\cite{2011rich} for variable star classification, iteratively reweights the training examples to increasingly focus on difficult-to-classify sources.  Random Forest,  which was used by multiple entrants in the Supernova Photometric Classification Challenge \cite{2010PASP..122.1415K} and by our group \cite{2011rich} for variable star classification, builds de-correlated trees by choosing a different random subset of features for each split in the tree-building process.  In Richards et al.~\cite{2011rich}, we found that random forest was the optimal method for a multi-class variable star problem in terms of error rate (Figure \ref{fig:misclass}).

\begin{figure}
\begin{center}
\includegraphics[angle=90,width=5.0in]{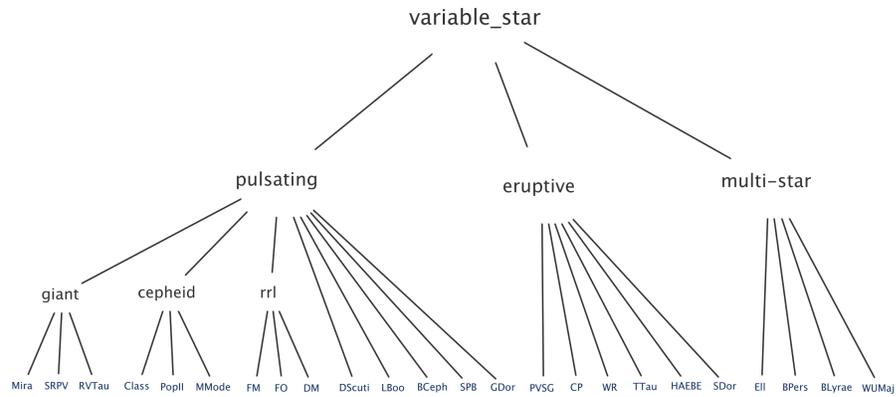}
\end{center}
\caption[Classification hierarchy of variable stars]{Variable star classification hierarchy for the problem considered in \cite{2011rich}.  This structure can be used in a hierarchical classifier to yield improved results.  The hierarchy is constructed based on knowledge of the physical processes and phenomenology  of variable stars.  At the top level, the sources split into three major categories: pulsating, eruptive, and multi-star systems.  \label{fig:classhierarchy}}
\end{figure}

In time-domain classification problems, we often have a well-established hierarchical taxonomy of classes, such as the variable star taxonomy in Figure \ref{fig:classhierarchy}.  Incorporating a known class hierarchy into a classification engine is a research field that has received much recent attention in the machine learning literature (e.g., \cite{2010sill}).  Several attempts for {\bf hierarchical classification} have been made in variable star problems.  Debosscher et al.~\cite{2009A&A...506..519D} use a 2-stage Gaussian mixture classifier, first classifying binaries versus non-binaries, while Blomme et al.~\cite{2010ApJ...713L.204B} use a multi-stage hierarchical taxonomy.  In Richards et al.~\cite{2011rich}, we use two methods for hierarchical classification, both using random forest and the taxonomy in Figure \ref{fig:classhierarchy}.

Finally, no discussion of supervised classification would be complete without mentioning the hugely-popular method {\bf Artificial Neural Networks} (ANN).  Though there are several versions of ANN, in their simplest form they are non-linear regression models that predict class as a non-linear function of linear combinations of the input features.  Drawbacks to ANN are their computational difficulty (e.g., there are many local optima) and lack of interpretability, and for these reasons they have lost popularity in the statistics literature.  However, they have enjoyed much success and widespread use in astronomy.  In time-domain astronomy, ANNs have been used by for variable star classification \cite{2005AJ....130...84F,2006A&A...446..395S,2007A&A...475.1159D} and by one team in the SN Classification Challenge (though the team's ANN entry fared much worse than their random forest entry, using the same set of features) .

\subsubsection{Unsupervised \& Semi-Supervised Approaches}

Unsupervised classification (statistical clustering) methods attempt to find $k$ clusters of sources in feature space.  These methods do not rely on any previously-known class labels, and instead look for natural groupings in the data.  After clusters are detected, labels or other significance can be affixed to them.  In time-domain studies, these methods are useful for explorative studies, for instance to discover the number of statistically-distinct classes in the data or to discover outliers and anomalous groups.  In the absence of confident training labels, an unsupervised study is a powerful way to characterize the distributions in the data to ultimately determine labels and build a predictive model using supervised classification.

In time-domain astronomy, the most popular clustering method is {\bf Gaussian Mixture Modeling}.  This method fits a parametric mixture of Gaussian distributions to the data by maximum likelihood via the expectation-maximization (EM) algorithm.  A penalized likelihood or Bayesian approach can be used to estimate the number of clusters present in the data.  The {\tt Autoclass} method \cite{1996chee} is a Bayesian mixture model clustering method that was used by Eyer \& Blake~\cite{2005MNRAS.358...30E} to cluster ASAS variable stars.  Sarro et al.~\cite{2009A&A...494..739S} use another variant of Gaussian Mixture Modeling to cluster a large database of variable stars.

{\bf Self-Organizing Maps} (SOM) is another popular unsupervised learning method in time-domain astronomy.  This method aims to map the high-dimensional feature vectors down to a discretized two-dimensional coordinate plane for easy visualization.  SOM is the unsupervised analog of ANN that uses a neighborhood function to preserve the topology of the input feature space.  This method has been used previously \cite{2004bret,2008AIPC.1082..201W} to obtain two-dimensional parametrization of astronomical light curves.  In those studies, SOM was performed prior to visual analysis of the labeled sources in this space.  This class of approach, where available class labels are ignored to obtain a simple parametrization of the light curve features and subsequently used in a learning step, is called \emph{semi-supervised learning}.  The advantage to this technique is that, if the relevant class information is preserved by the unsupervised step, then supervised classification will be easier in the reduced space.  Semi-supervised classification permeates the time-domain astronomy literature.  In addition to the afore-mentioned SOM studies, other authors have used PCA \cite{2007arXiv0712.2898W, 2009A&A...507.1729D} and diffusion map \cite{2011richSN} to parametrize time-variable sources prior to classification.  Of these studies, only Richards et al.~\cite{2011richSN} used a rigorous statistical classifier.

\section{Future Challenges}

For any finite collection of photons, our knowledge of the true flux is inherently uncertain. This basic phenomenological uncertainty belies an even greater uncertainty in the physical origin of what we think we are witnessing. As such, any classification scheme of a given variable or transient source must be  inherently probabilistic in nature. We have outlined how---with an emerging influence of the  machine-learning literature---we can gain traction on the probabilistic classification challenge. 
Calibrating (and validating) the output probabilities from  machine-learning frameworks is still a nascent endeavor. 

Feature generation is obviously a key ingredient to classification and we have presented evidence that random forest classifiers are particularly useful at using features that are most relevant to classification and skirting the problem of large covariance between features. On the positive side, this frees us from having to create a small set of perfectly tuned features. However, how do we know when we have exhausted the range of reasonable feature space for classification? Our suspicion is that expert knowledge has already imbued the feature creation process with much of the knowledge implicitly needed for classification: we know for instance that phase offset between the first and second most dominant periods can be a powerful way to distinguish two closely related classes of pulsational variables. There may be information-theoretic (and feature-agnostic) answers to this question, which might be attacked with some genetic programming framework.

On statistical grounds, implicit in the feature generation procedure is that the distribution of features (and their covariances) on the training set will be similar to the set of instances of sources we wish to classify. A gross mismatch of the characteristics of these two sets is likely to be  a significant problem for the robustness of the classification statements.  No study to date has looked at how we can use the knowledge gleaned from one survey and apply that to classification in another. For instance, if a classifier is blindly trained on one survey to classify objects from another, then it will achieve sub-optimal results by not considering differences in feature distribution between the surveys.  Ideas from statistics, such as importance sampling, can be exploited to account for these differences.

As these very basic algorithmic questions are addressed, the computational implications, using events from real surveys, will have to be understood. Is feature creation and the application of an existing machine-learned framework fast enough for a given data stream? How can loss functions be embedded in computational choices at the feature and the labeling levels? For streaming surveys, how often should the learning model be updated with newly classified examples from the survey itself? What are the roles of massively parallel hardware (e.g. graphical processing units) in feature generation, learning, and classification?

Astronomical datasets have always presented novel algorithmic, computational, and statistical challenges. With classification based on noisy and sometimes-spurious data,  the forefront of all of these endeavors  is already being stretched. As astronomers, expanding the machine-learning literature is a means to an end---if not just a way to keep our heads above water---building a vital set of tools for the exploration of the vast and mysterious dynamic universe.

\bigskip

{\it The authors acknowledge the generous support of a CDI grant (\#0941742) from the National Science Foundation. We thank Nat Butler and Dan Starr for helpful conversations. J.S.B. thanks those at the Universitat de Barcelona (Spain) for accommodating him in the astronomy department in late 2010, where much of this chapter was written.}

\bibliographystyle{plain}
\bibliography{biblio,journals_apj}

\printindex

\end{document}